\documentclass[%
 reprint,
 amsmath,amssymb,
 aps, prl
]{revtex4-2}

\usepackage{graphicx}
\usepackage{bm}
\usepackage{hyperref}
\usepackage{xcolor}
\usepackage{soul}

\begin{document}

\preprint{APS/123-QED}

\title{Sachs-Wolfe effect as a smoking gun for cosmological gravitational wave backgrounds}

\author{Giorgio Mentasti}
 \email{giorgio@apc.in2p3.fr}
\author{L\'eon Vidal}%
 \email{vidal@apc.in2p3.fr}
\author{Quentin Baghi}%
\affiliation{Universit\'e Paris Cit\'e, CNRS, Astroparticule et Cosmologie, F-75013 Paris, France
}%
\author{Carlo~R.~Contaldi}
\affiliation{
 Department of Physics, Imperial College London, SW7 2AZ, London, United Kingdom
}%

\date{\today}

\begin{abstract}
The Sachs-Wolfe (SW) effect, arising from large-scale structures in the universe, modifies the frequencies of gravitational waves (GWs) sourced by a cosmological background. We show that for backgrounds with $\Omega_{\rm GW}\gtrsim 10^{-10}$, this effect imprints anisotropies and spectral distortions that can be detectable with a network of space-based interferometers (such as LISA + Taiji) and, if not taken into account, may bias the estimate of the theoretical model of the GW background.  The effect is particularly enhanced in the high-frequency end of the spectrum. The SW-induced anisotropies and spectral distortions present in a GW background sourced at primordial times will correlate with the SW signature present in the CMB. Any detection of a cross-correlation between the GW anisotropies and the CMB at large scales is therefore a smoking gun for confirming the primordial nature of the background.
\end{abstract}

\maketitle


\section{Introduction}
The detection of a stochastic gravitational wave background (SGWB) in the millihertz band promises to unlock a new observational window into the early universe. As the first space-based gravitational wave interferometers, LISA \cite{Baker:2019nia} and Taiji \cite{doi:10.1142/S0217751X2050075X,Liang:2021bde} will probe cosmological sources—such as first-order phase transitions (PTs)\cite{Caprini:2024hue}, primordial black hole (PBH) \cite{Sasaki:2018dmp} formation and cosmic strings \cite{Damour:2004kw}—that are inaccessible to ground-based detectors. These sources generate relic gravitational wave (GW) backgrounds with distinctive spectral features, encoding fundamental physics at energy scales far beyond terrestrial colliders. Future space-based detectors’ ability to measure the spectral shape, amplitude, and anisotropies of an SGWB makes them a unique tool for testing early-universe scenarios and constraining new physics models.

Stochastic GW backgrounds are challenging to detect with a single observatory, because they share common characteristics with instrumental noise. Works studying detection with LISA alone usually rely on either parametrized spectral templates for the signal \cite{christensenStochasticGravitationalWave2019,baghiUncoveringGravitationalwaveBackgrounds2023,santiniFlexibleGPUacceleratedApproach2025a} or precise knowledge of noise spectral characteristics \cite{capriniReconstructingSpectralShape2019,flaugerImprovedReconstructionStochastic2021,pozzoliWeaklyParametricApproach2024,kumeAssessingImpactUnequal2025}, or both \cite{cornishDetectingStochasticGravitational2001,boileauSpectralSeparationStochastic2021a,criswellTemplatedAnisotropicAnalyses2024,buscicchioFirstYearLISA2025}. Flying two detectors simultaneously would break the degeneracy between noise and GW backgrounds, increasing the level of confidence required for potential scientific discoveries \cite{wangAbilityLISATaiji2024,chengDetectionStochasticGravitational2025,liangUnveilingMulticomponentStochastic2025}. Still, distinguishing between backgrounds resulting from a population of compact binary objects and cosmological signals from the early universe remains delicate. Nonetheless, any characteristic sky anisotropy of the background could be leveraged to make this distinction.

SGWBs of cosmological origin may exhibit different levels of anisotropies acquired at the production epoch or during their propagation throughout the universe. In particular, the SW effect \cite{1967ApJ...147...73S,Hu:1994jd}, imprints anisotropies in any incoming radiation due to inhomogeneities in the local potential at emission. If the emission surface is sufficiently early, this effect introduces a unique anisotropic pattern and spectral distortion that cannot be mimicked by astrophysical backgrounds or noise. This makes the SW effect a smoking gun for cosmological GW backgrounds.
In this work, we model the perturbation induced by the SW effect on a GW spectrum of cosmological origin. We forecast the signal-to-noise ratio (SNR) of the difference between the model that accounts for the SW effect and the one that does not account for it. As another figure of merit for assessing the detectability of this effect, we quantify the potential bias in parameter estimates if this effect is neglected and compare it to the forecast errors in their measurement. We assess the potential for using the SW effect to identify and characterise a cosmological SGWB with LISA and a network comprising LISA \cite{Baker:2019nia} and Taiji \cite{doi:10.1142/S0217751X2050075X,Liang:2021bde}.

\section{Sachs--Wolfe Effect for a Stochastic Gravitational Wave Background}

Any massless radiation field propagating through an inhomogeneous universe experiences gravitational redshift due to metric perturbations. This applies equally to photons and gravitons, leading to a shift in their observed frequencies. The effect is perturbative on large angular scales where curvature perturbations are small. At linear order in the metric perturbations, the effect can be written in terms of the potentials at the source and observer locations and the integrated evolution of potentials along the line of sight, the latter known as the integrated Sachs-Wolfe (ISW) effect\cite{1967ApJ...147...73S,1985SvAL...11..271K}. Assuming the incoming radiation is observed at a single location, the endpoint boundary term is unobservable, and only the SW effect from the emission boundary is observable.

It is convenient to parameterise the frequency shift along a direction $\hat n$ as $f_{\rm obs}(\hat n) = f_{\rm em}(\hat n)\,(1+\Gamma(\hat n))$ where $\Gamma(\hat n)$ maps the frequency distortion as a function of direction on the sky. For scalar perturbations, the line-of-sight solution for $\Gamma$ is \cite{1996ApJ...469..437S,Dodelson:2003ft,Contaldi:2016koz,Cusin:2017fwz,Bartolo:2019oiq}
\begin{equation}
\Gamma(\hat n)
=
\left[\Psi\right]_{\rm em}^{\rm obs}
+
\int_{\eta_s}^{\eta_0} d\eta\,\left(\dot\Phi+\dot\Psi\right),
\label{eq:Gamma_def}
\end{equation}
where $\eta_s$ is the conformal time of emission, $\eta_0$ is today, $\Psi$ and $\Phi$ are the Newtonian gauge potentials describing the perturbations around the Friedmann–Robertson–Walker spacetime, and the overdots denote derivatives with respect to conformal time. The first term corresponds to the SW effect, while the integral term corresponds to the ISW effect.

We consider an SGWB generated instantaneously at conformal time $\eta_s$, with an isotropic spectral distribution function $\bar\Delta_{\rm gw}(f)$. In the absence of propagation effects, the observed background remains isotropic. However, the frequency remapping induced by $\Gamma(\hat n)$ leads to an observed spectrum distribution \cite{Mentasti:2025ywl}
\begin{equation}
\Delta_{\rm gw}^{\rm obs}(f,\hat n)
=\bar\Delta_{\rm gw}\!\left(f[1+\Gamma(\hat n)]\right).
\label{eq:exact_remap}
\end{equation}
Eq.~\eqref{eq:exact_remap} highlights that the effect of gravitational redshift is a remapping in frequency space. The expression is usually Taylor expanded; however, this expansion is not generally valid for SGWB spectra with sharp features or spikes, such as those arising from primordial black hole formation or second-order induced backgrounds. In such cases, the exact mapping Eq.~\eqref{eq:exact_remap} must be used, as emphasised in Ref.~\cite{Mentasti:2025ywl}, where it was shown that a perturbative expansion fails in the presence of large spectral gradients.

Even if the primordial SGWB is perfectly isotropic, Eq.~\eqref{eq:exact_remap} implies that anisotropies are generated by the directional dependence of $\Gamma(\hat n)$. Expanding the anisotropic redshifting field in spherical harmonics,
\begin{align}\label{eq:map_def}
\Gamma(\hat n) = \sum_{\ell m} a_{\ell m} Y^{\,}_{\ell m}(\hat n)\,,
\end{align}
we define the angular power spectrum, which can be calculated using the standard line-of-sight method \cite{1996ApJ...469..437S},  yielding
\begin{equation}\label{eq:ClGamma}
C_\ell \equiv \langle |a_{\ell m}|^2 \rangle = 4\pi \int \frac{dk}{k}\,\mathcal P_{\mathcal R}(k)
\,|\Delta_\ell(k,\eta_0)|^2,
\end{equation}
where $k$ is the wavenumber and $\mathcal{P}_\mathcal{R}(k)$ is the spectrum of primordial curvature perturbations and the angular transfer function are defined as
\begin{align}\label{eq:LOS}
   \Delta_\ell(k,\eta_0) &= \Psi(k,\eta_s)j_\ell(k[\eta_0-\eta_s])+\nonumber\\
&\int_{\eta_s}^{\eta_0} d\eta\,
(\dot\Phi+\dot\Psi)\,
j_\ell\bigl(k[\eta_0-\eta]\bigr).
\end{align}
Note that we have assumed the SGWB has no intrinsic perturbations and that it is created instantaneously.

We evaluate Eq.~\eqref{eq:ClGamma} numerically using a fiducial $\Lambda$CDM cosmology. The time evolution of the metric perturbations is obtained using \texttt{CAMB} \cite{Lewis:1999bs}, from which we extract the Weyl potential $\phi(k,\eta) = (\Phi(k,\eta)+\Psi(k,\eta))/2$ as a function of $k$ and $\eta$ (we assume anisotropic stresses are negligible such that $\phi=\Psi$ for simplicity). The line-of-sight integral is then computed explicitly from the source conformal time $\eta_s$ to the present, and the angular power spectrum is obtained by integrating over the primordial curvature power spectrum.

Although the SGWB is generated at conformal times well before recombination, the comoving distance to the source surface is very close to that of the CMB last-scattering surface, yielding nearly identical angular transfer functions that probe similar wavenumbers. Since the horizon at the time of SGWB generation is much smaller than at recombination, the corresponding SW plateau extends to significantly higher multipoles. However, in practice, the limited angular resolution of gravitational wave observations (see, for example, \cite{Mentasti:2023uyi}) restricts access to these small-scale anisotropies, which means that observations of the effect introduced here are insensitive to the exact value of $\eta_s$.

As a result, SGWB anisotropies sourced by the SW effect are highly correlated with those of the CMB \cite{Cai:2024dya,Bartolo:2019yeu}, since both arise from the same underlying metric perturbations. The relative contribution of the SW term is larger for the SGWB because the CMB temperature anisotropy includes an intrinsic photon density perturbation that is partially anticorrelated with the gravitational redshift, leading to a suppression of the net SW signal. The ISW effect is therefore even more subdominant compared to the CMB case and constitutes a nuisance foreground since it is present in any background, including those sourced at astrophysical distances.

\begin{figure}[!t]
	\centering
	\includegraphics[width=\linewidth,clip]{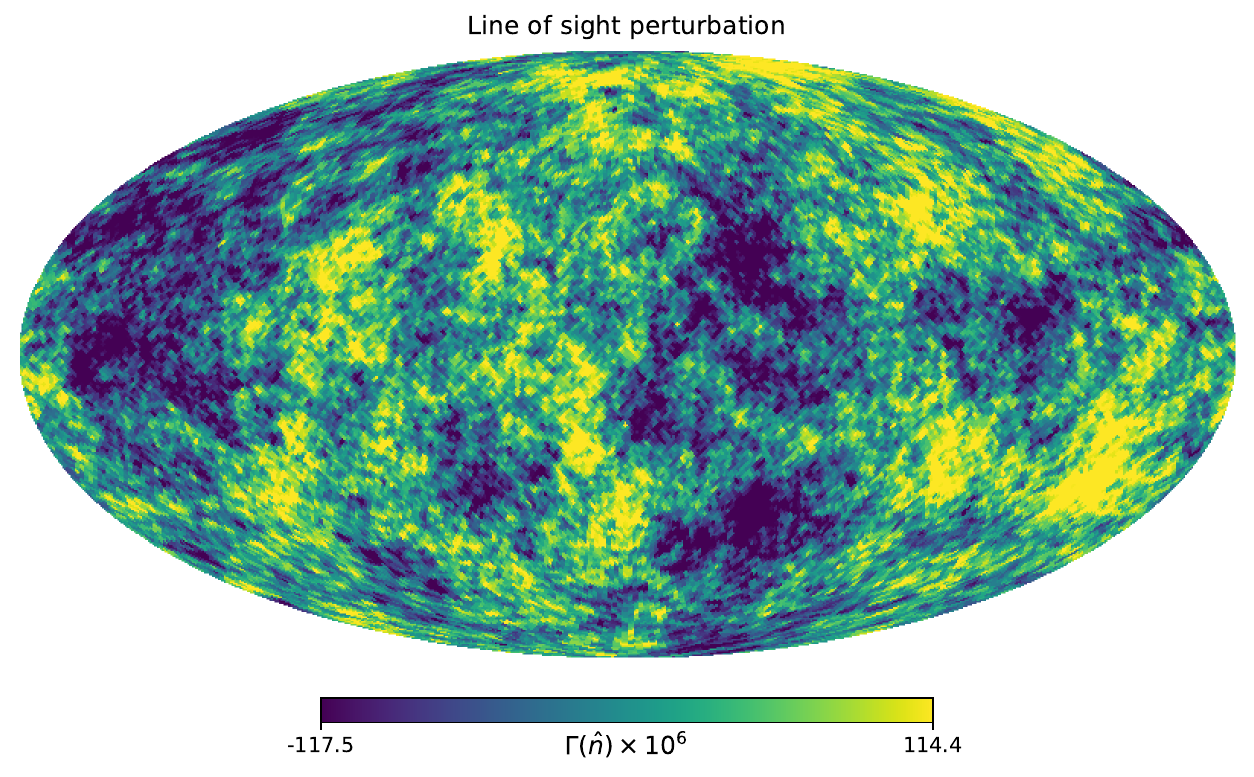}
	\caption{Simulated map of the SW anisotropies defined in Eq.~\eqref{eq:map_def}. In practice, only the largest angular scales present in this map are accessible using GW observations.}
	\label{fig:map}
\end{figure}

At low multipoles, reconstructed maps of the SW contribution can be obtained, for example, via deconvolution of the source angular transfer functions from CMB maps \cite{2001PhRvD..63f3002K,Komatsu:2003iq}, a technique commonly used in reconstructing maps of the primordial fluctuations to constrain primordial non-Gaussianity \cite{Planck:2019kim,Bravo:2025csu,Zhao:2024gan}.  For simplicity, here, we estimate the signal-to-noise ratio using simulated maps generated from the $C_\ell$ spectra derived above.  As the signal is dominated by the lowest multipoles, where the CMB-derived template is signal-dominated, this yields a good approximation to the true signal-to-noise ratio. Fig.~\ref{fig:map} shows a simulated map of the anisotropies $\Gamma(\hat n)$ obtained from Eq.~\eqref{eq:ClGamma}.

\section{Analysis}

Let us assume that we have a GW background that is originally isotropic and stationary, with power spectrum $S_h^0(f)$. At observation, the amplitude of the GW frequency domain $\tilde h_\lambda$ is anisotropic and spectrally distorted with respect to emission\cite{Mentasti:2025ywl}
\begin{align}\label{eq:SGWB_def}
&\langle \tilde h_\lambda(f,\hat n) \tilde h^*_{\lambda'}(f',\hat n')\rangle = S_h(f,\hat n)\delta(f-f')\delta(\hat n-\hat n')\delta_{\lambda\lambda'}\,,\nonumber\\
&S_h(f,\hat n)=S_h^0\left(f\,\Gamma(\hat n)\right)\Gamma^{-1}(\hat n)\,,
\end{align}
where $\lambda$ indicates the polarisation mode. From the explicit computation of the \eqref{eq:ClGamma} we find that the $C_{\ell}$ coefficients of the spherical harmonics decomposition of the SW induced distortion map $\Gamma(\hat n)$ defined in \eqref{eq:map_def} are
\begin{align}
\langle a_{\ell m}a^*_{\ell' m'}\rangle &= C_{\ell} \delta_{\ell \ell'}\delta_{mm'}\simeq 1.2\times10^{-9}\frac{2\pi\,\delta_{\ell \ell'}\delta_{mm'}}{\ell(\ell+1)}\,.
\end{align}
We produce this map by drawing a realisation of the $a_{\ell m}$ coefficients with the {\tt HEALPY} software \cite{Zonca2019healpy}. As explained in the previous sections, an estimate of the real SW map can be obtained using existing CMB surveys, and in the following, we assume sufficiently precise knowledge of it.
In fact, we decided not to use that real map, since doing so would bring further technical, survey-dependent complications in the analysis that go beyond the scope of our work. Our simulated map is a faithful representation of the real one: we verified this by drawing different realisations of the $a_{\ell m}$ coefficients and found that the results depend weakly on the individual realisations.

\begin{figure}[!t]
	\centering
	\includegraphics[width=\linewidth,clip]{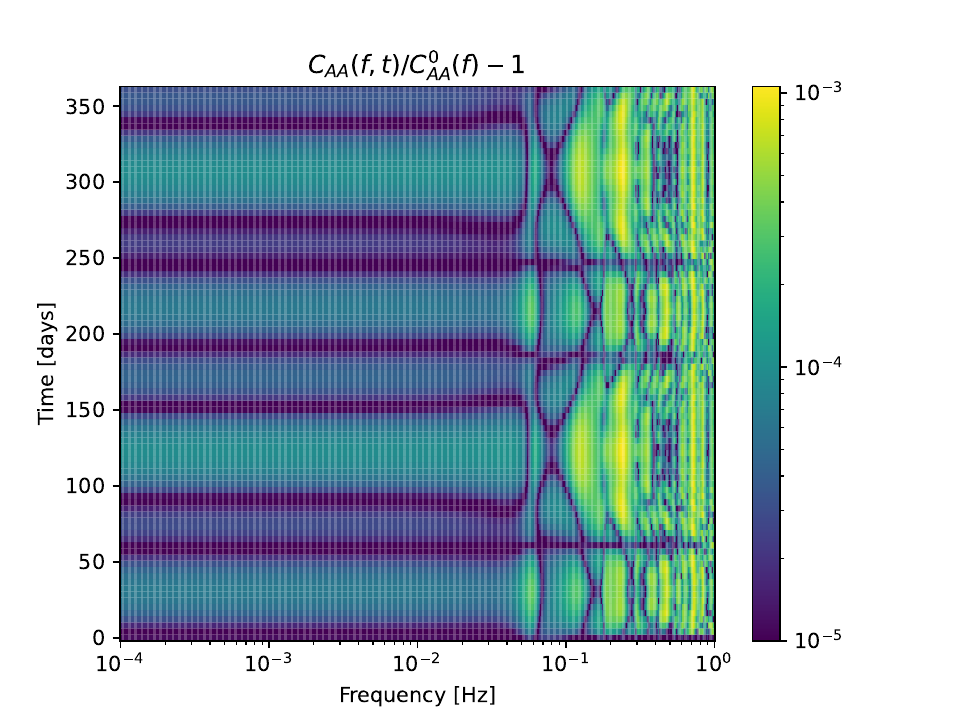}
x	\caption{The relative difference in the response functions $C_{ij}(f,t)$ and $C_{ij}^0(f)$ defined in eq. \eqref{eq:response_correct} and \eqref{eq:response_wrong} as a function of time and frequency assuming a power law model $\Omega_{\rm GW}(f)=\Omega_0(f/f_0)^\alpha$ (we set a fiducial value of $\bar\alpha=0$ and the pivot frequency at $f_0=1$mHz). Here we consider the self correlator in the AET basis $(i,j)=(A,A)$ for LISA as an example.}
	\label{fig:response}
\end{figure}

\begin{figure}[!t]
	\centering
	\includegraphics[width=\linewidth,clip]{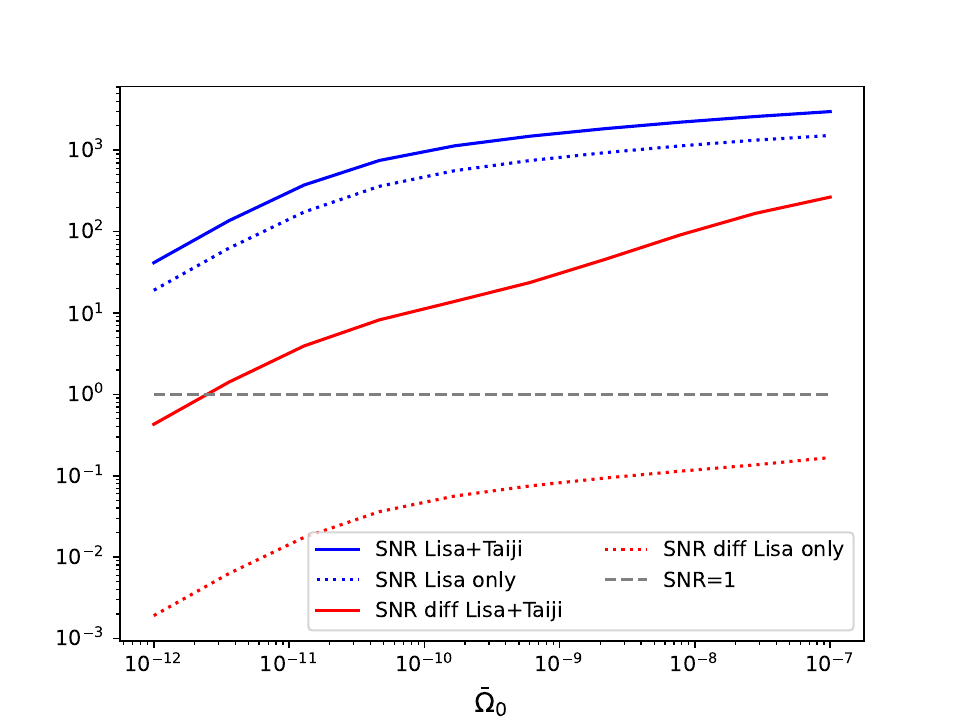}
	\caption{The SNR of the residual between the wrong and the correct model and the absolute value of the SNR of the unperturbed model when measured with LISA and LISA + Taiji. A power law model $\Omega_{\rm GW}(f)=\Omega_0(f/f_0)^\alpha$ (we set a fiducial value of $\bar\alpha=0$), as a function of the fiducial value $\bar\Omega_0$. A time of observation of $T_{\rm obs}=2$yr is assumed.}
	\label{fig:snrs}
\end{figure}

\begin{figure}[!t]
	\centering
	\includegraphics[width=\linewidth,clip]{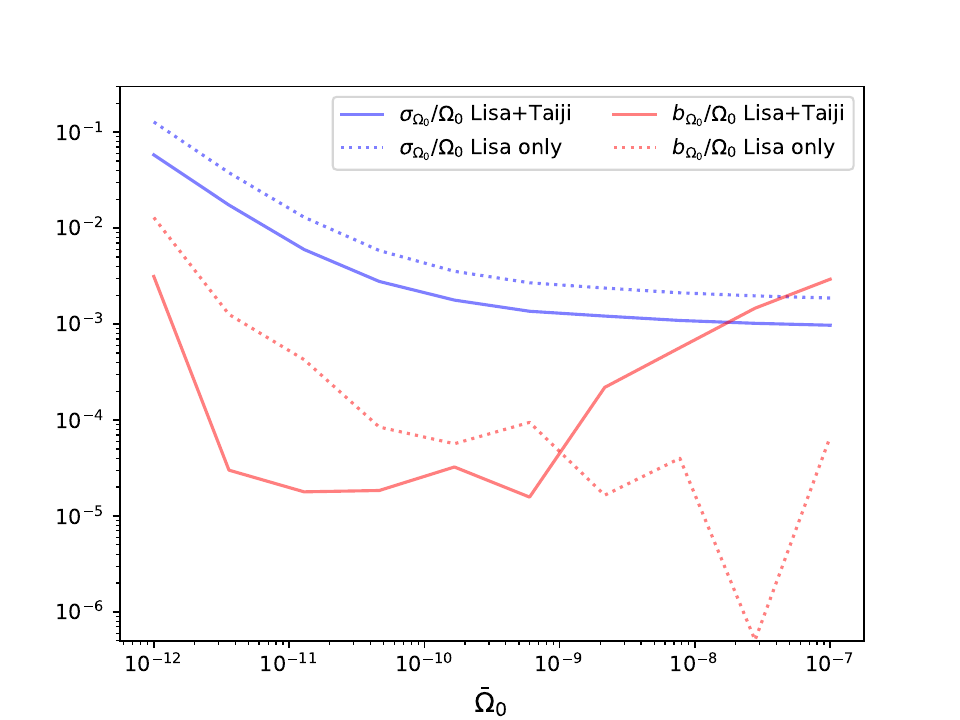}
	\caption{The forecast bias and marginal errors on $\Omega_0$ while inferring with LISA alone and LISA + Taiji the parameters $\Omega_0$ and $\alpha$ for a power law model $\Omega_{\rm GW}(f)=\Omega_0(f/f_0)^\alpha$ (we set a fiducial value of $\bar\alpha=0$ and the pivot frequency at $f_0=1$mHz), as a function of the fiducial value $\bar\Omega_0$. The values are obtained after marginalising the posterior distribution over $\alpha$. A time of observation of $T_{\rm obs}=2$yr is assumed.}
	\label{fig:lines_omega}
\end{figure}

\begin{figure}[!t]
	\centering
	\includegraphics[width=\linewidth,clip]{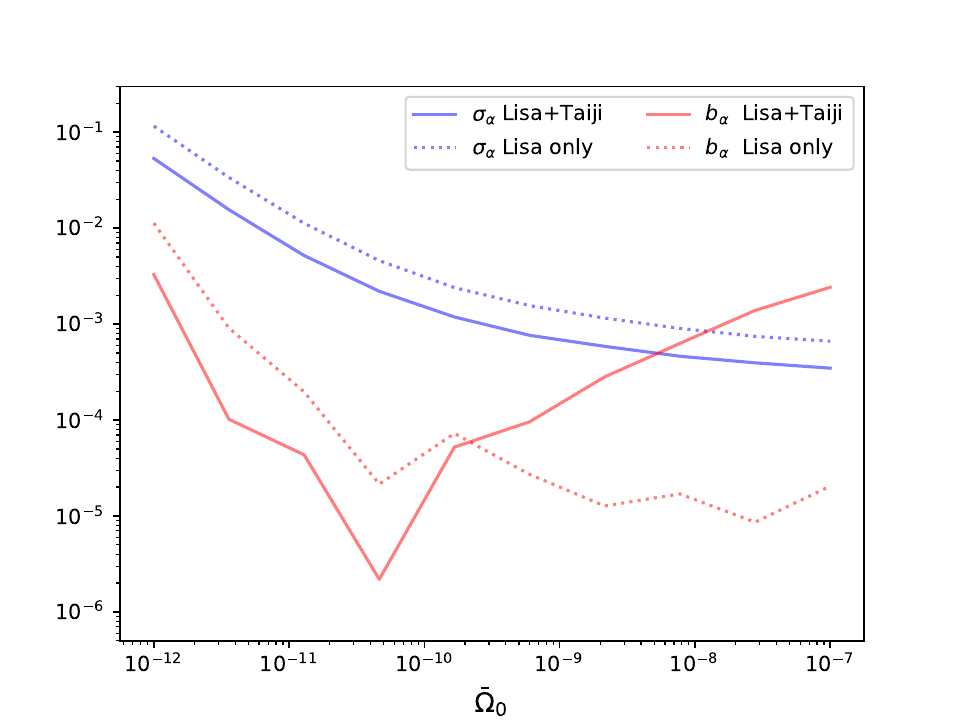}
	\caption{The forecast bias and marginal errors on $\alpha$ while inferring with LISA alone and LISA + Taiji the parameters $\Omega_0$ and $\alpha$ for a power law model $\Omega_{\rm GW}(f)=\Omega_0(f/f_0)^\alpha$ (we set a fiducial value of $\bar\alpha=0$ and the pivot frequency at $f_0=1$mHz), as a function of the fiducial value $\bar\Omega_0$. The values are obtained after marginalising the posterior distribution over $\Omega_0$. A time of observation of $T_{\rm obs}=2$yr is assumed.}
	\label{fig:lines_alpha}
\end{figure}

The observables for LISA (and similarly for LISA + Taiji) are the set of correlators between the data streams of the surveys, which we denote by the indices $i,j=1\dots3$ for LISA time-delay interferometry AET channels~\cite{princeLISAOptimalSensitivity2002} and $i,j=4\dots6$ for Taiji AET channels. Assuming perfect knowledge of the noise statistics, we debias the correlators by subtracting the noise PSDs. We assume Taiji has the same arm length and instrumental noise as LISA, while we model the orbits' geometry as in \cite{Mentasti:2023uyi}. In the time-frequency domain \cite{LISACosmologyWorkingGroup:2022kbp,Mentasti:2023uyi}, we can define the forecast expected value of the correlators $\langle C_{ij}(f,t;\theta)\rangle$ between TDI channels, where $\theta$ represents the set of theoretical parameters used to model the background power spectrum. The correlators are written in terms of the GW spectrum resulting from the superposition of the waves coming from all the sky directions:
\begin{align}\label{eq:response_correct}
\langle C_{ij}(f,t;\theta)\rangle &=\int d^2\hat n\,S_h(f,\hat n;\theta)\mathcal{R}_{ij}(f,\hat n,t)\,.
\end{align}
On the other hand, if one wrongly assumes the unperturbed model (i.e. without taking into account the SW effect) we would naively assume a form
\begin{align}\label{eq:response_wrong}
\langle C_{ij}^0(f;\theta)\rangle&=S_h^0(f;\theta)\int d^2\hat n\,\mathcal{R}_{ij}(f,\hat n,t)\,,
\end{align}
where the response functions $\mathcal{R}_{ij}(f,\hat n,t)$ are defined following the conventions of \cite{LISACosmologyWorkingGroup:2022kbp}, before the angular integration. The difference between the two models \eqref{eq:response_correct} and \eqref{eq:response_wrong} is qualitatively illustrated by Fig. \ref{fig:response}.
The (log)likelihood assuming the correct template (that includes the SW effect) and the one that does not do so are defined, respectively, as \cite{Mentasti:2023uyi}
\begin{align}\label{eq:likelihood}
&\mathcal{L}_{\rm SW}(\theta)=-\frac{1}{2}\sum_{ij}\int df\int  dt \frac{\left|\langle C_{ij}(f,t;\theta)\rangle -\langle C_{ij}(f,t;\bar\theta)\rangle \right|^2}{D_{ij}(f,t;\bar\theta)}\,,
\end{align}
and
\begin{align}\label{eq:likelihood_biased}
&\mathcal{L}_{\rm no SW}(\theta)=-\frac{1}{2}\sum_{ij}\int df\int dt \frac{\left|\langle C_{ij}^0(f;\theta)\rangle -\langle C_{ij}(f,t;\bar\theta)\rangle\right|^2}{D_{ij}(f,t;\bar\theta)}\,,
\end{align}
with
\begin{align}
&D_{ij}(f,t;\bar\theta)=(\langle C_{ii}^{0*}(f,t;\bar\theta)\rangle +N_i(f))(\langle C_{jj}^0(f,t;\bar\theta)\rangle +N_{j}(f))\nonumber\\
&+(\langle C_{ij}^{0*}(f,t;\bar\theta)\rangle +\delta_{ij}N_i(f))(\langle C_{ji}^0(f,t;\bar\theta)\rangle +\delta_{ij}N_{j}(f))\,,
\end{align}
where $\bar\theta$ is the fiducial value of the set of theoretical parameters and $N_{i}(f)$ are the noise PSD functions for the TDI channels $i$. Both $N_i(f)$ and $\langle C_{ij}(f,t)\rangle $ can be computed in any TDI combination, being careful to have the two consistent with each other. For the sake of our work, we chose the first generation of TDI, but the treatment does not change at all for any other combination \cite{LISACosmologyWorkingGroup:2022kbp,Mentasti:2023uyi}. The first likelihood function in Eq.~\eqref{eq:likelihood} is unbiased by definition: the best fit value $\theta_{\rm SW}$ is equal to $\bar\theta$. On the other hand, the best fit value obtained by using the wrong model in Eq.~\eqref{eq:likelihood_biased} leads to a bias $b_\theta=\theta_{\rm no SW}-\bar\theta$.
Furthermore, from the curvature of the likelihood function~\footnote{At first order, one can equivalently use the correct or the wrong likelihood function to estimate the forecast error.} we can estimate the error $\sigma_\theta$ in the measurement of the theoretical parameters $\theta$.
We compute $b_\theta$ and $\sigma_\theta$ via the evaluation of the two likelihoods for a PSD with almost-flat energy density $\Omega(f)=\Omega_0(f/f_0)^\alpha$, where we assume a fiducial value of $\bar\alpha=0$ and we set $f_0=1$ mHz.
We then sample the likelihood distribution via the Markov chain Monte Carlo (MCMC) based sampler \textit{emcee} \cite{Foreman_Mackey_2013}.
In figures \ref{fig:lines_omega} and \ref{fig:lines_alpha}, we show how the bias and measurement error on the parameters $\theta=\{\Omega_0,\alpha\}$ scale given the fiducial value of the energy density $\bar\Omega_0$. We see that while LISA alone will always exhibit a bias much smaller than the measurement error when using the wrong likelihood, this will be different for a network comprising LISA and Taiji.
As another figure of merit, we compute the signal-to-noise ratio (SNR) of the difference between the correct model and the one that does not take into account the SW effect:
\begin{align}\label{eq:snr}
{\rm SNR}_{\rm diff}=\sqrt{\sum_{ij}\int df\int dt \frac{\left|\langle C_{ij}(f,t,\bar\theta)\rangle -\langle C_{ij}^0(f,\bar\theta)\rangle \right|^2}{D_{ij}(f,t,\bar\theta)}}\,.
\end{align}
This can also be considered as the SNR of the residual that one may still have after fitting and subtracting the wrong model from the stochastic background correlator estimate $\langle C_{ij}(f,t)\rangle $.
We compute such an SNR of the difference for the two detector configurations (LISA alone and LISA + Taiji), and we compare it with the absolute value of the SNR of the background~\footnote{Also in this case, to compute the absolute value of the SNR, one can equivalently use the wrong or the correct model, as they are equivalent at first order.}. In figure~\ref{fig:snrs} we show how these SNR values scale with the fiducial value of the energy density $\Omega_0$ for the same quasi-flat power law model.

Figures \ref{fig:snrs}, \ref{fig:lines_omega}, and \ref{fig:lines_alpha} show that LISA alone will not be able to observe the imprint of the SW effect on the timescale of the mission: the very low relative amplitude of the anisotropies and LISA's poor angular resolution are such that this effect is completely masked by the instrumental noise.
On the other hand, the combination of LISA and Taiji gives a better prospect for detecting this effect as a smoking gun for a cosmological origin of the GW background. The fundamental reason why an additional instrument brings about such a large improvement is the fact that the anisotropic response functions show a peak in frequency that depends on the baseline between the instruments \cite{LISACosmologyWorkingGroup:2022kbp,Mentasti:2023uyi}, in such a way that relatively smaller angles are much better resolved by large baselines.
For the same reason, we find that LISA and LISA + Taiji are even more sensitive to this effect when considering blue-tilted power spectra. In order not to overcomplicate the display of our results, we decide to show only the flat $\Omega_{\rm GW}$ case (which is also more conservative).
As can be seen in figure \ref{fig:snrs}, the imprint of the SW effect can already induce a significant effect for $\Omega_0 > 10^{-10}$, while there is no measurable bias in the parameter inference for $\Omega_0<10^{-8}$, as in figures \ref{fig:lines_omega} and \ref{fig:lines_alpha}.
The reason is the fact that the SNR of the difference between the observed background and the unperturbed model (defined in Eq.~\eqref{eq:snr}) accumulates the effect arising from the integration of the absolute value of the fluctuations in frequency and time.
On the other hand, the joint fit of $\Omega_0$ and $\alpha$ interpolates these features. When the deviations from the unperturbed model become large enough, the fit with the unperturbed model fails, and biases appear, as seen in figures \ref{fig:lines_omega} and \ref{fig:lines_alpha}.

By cross-correlating LISA and Taiji, one can disentangle instrumental noise from stochastic backgrounds much more effectively than with LISA alone.
However, there is no universal way to distinguish a cosmological background from one of astrophysical origin: to our knowledge, the effect proposed in the present work is the first to provide a model-independent and potentially observable signature of a cosmological origin.
Lastly, one may argue that the cosmological background may present anisotropies at production. For these kinds of backgrounds, one may apply the same formalism used in the present work to study the intrinsic anisotropies as well as those induced by the SW effect. In this case, including this effect is not only useful in order to prove the cosmological origin of such a background, but it will be essential in order to prove that the observed anisotropies are intrinsic (in a similar way to what has been done in CMB experiments).

\section{Conclusions}
Our results show that the SW induces both angular and spectral distortions in an initially isotropic SGWB. Through directional redshifting induced by large-scale gravitational potentials, this effect imprints an angular and spectral modulation of the signal, whose pattern is fixed by the propagation history of the background.
In principle, a modulation due to the late ISW effect is also present, but this is subdominant. We find that this effect is one order of magnitude or smaller than the SW effect and therefore does not impact our analysis. 

Regarding the forecast sensitivity for detecting this effect, when LISA alone is considered, its imprint remains hidden by limited angular resolution and instrumental noise in the most natural scenarios of expected cosmological backgrounds. In contrast, a network made of two space-based interferometers (such as LISA and Taiji) provides a qualitatively different regime: the long inter-detector baseline enhances the anisotropic response and makes the SW residual potentially observable, reaching a potential SNR of the difference between perturbed and unperturbed models of $\mathrm{SNR}_{\rm diff}>10$ for backgrounds with fiducial amplitudes around $\Omega_0\gtrsim 10^{-10}$.
In this regime, the SW imprint is not only a discovery signature (distinguishing a GW background of primordial origin from any other foregrounds): it also becomes an essential part of the signal model required to interpret a cosmological SGWB. Neglecting it would leave coherent residuals in the data and could bias parameter inference, while its large-scale modulation pattern provides a way to distinguish a cosmological component from astrophysical backgrounds. Our results, therefore, show that a LISA + Taiji network would not only improve sensitivity to SGWBs but also enable access to a propagation signature of their primordial origin.
Lastly, it is important to note that the analysis presented here can be applied to other GW surveys. In particular, we believe that it is possible to exploit the SW effect in analysing the GW background for the next generation of ground-based observatories (such as the Einstein Telescope or the Cosmic Explorer) to disentangle the cosmological from the astrophysical backgrounds. We will tackle this analysis in future work.

\begin{acknowledgments}
GM acknowledges support from the ANR grant AAPG2024 PRC - GalaxyFIT. LV acknowledges support from AP Crossing Cutting Edges Program of Université Paris Cité.
\end{acknowledgments}

\nocite{*}

\bibliography{refs}

@article{Hu:1994jd,
    author = "Hu, Wayne and Sugiyama, Naoshi",
    title = "{Toward understanding CMB anisotropies and their implications}",
    eprint = "astro-ph/9411008",
    archivePrefix = "arXiv",
    reportNumber = "CFPA-TH-94-55, CFPA-94-TH-55, UTAP-193",
    doi = "10.1103/PhysRevD.51.2599",
    journal = "Phys. Rev. D",
    volume = "51",
    pages = "2599--2630",
    year = "1995"
}

@article{Planck:2019kim,
    author = "Akrami, Y. and others",
    collaboration = "Planck",
    title = "{Planck 2018 results. IX. Constraints on primordial non-Gaussianity}",
    eprint = "1905.05697",
    archivePrefix = "arXiv",
    primaryClass = "astro-ph.CO",
    doi = "10.1051/0004-6361/201935891",
    journal = "Astron. Astrophys.",
    volume = "641",
    pages = "A9",
    year = "2020"
}

@ARTICLE{2001PhRvD..63f3002K,
       author = {{Komatsu}, Eiichiro and {Spergel}, David N.},
        title = "{Acoustic signatures in the primary microwave background bispectrum}",
      journal = {\prd},
         year = 2001,
        month = mar,
       volume = {63},
       number = {6},
          eid = {063002},
        pages = {063002},
          doi = {10.1103/PhysRevD.63.063002},
archivePrefix = {arXiv},
       eprint = {astro-ph/0005036},
 primaryClass = {astro-ph},
       adsurl = {https://ui.adsabs.harvard.edu/abs/2001PhRvD..63f3002K}
}

@article{Komatsu:2003iq,
    author = "Komatsu, Eiichiro and Spergel, David N. and Wandelt, Benjamin D.",
    title = "{Measuring primordial non-Gaussianity in the cosmic microwave background}",
    eprint = "astro-ph/0305189",
    archivePrefix = "arXiv",
    doi = "10.1086/491724",
    journal = "Astrophys. J.",
    volume = "634",
    pages = "14--19",
    year = "2005"
}

@article{Lewis:1999bs,
    author = "Lewis, Antony and Challinor, Anthony and Lasenby, Anthony",
    title = "{Efficient computation of CMB anisotropies in closed FRW models}",
    journal = "Astrophys. J.",
    volume = "538",
    pages = "473--476",
    year = "2000",
    eprint = "astro-ph/9911177",
    archivePrefix = "arXiv",
    primaryClass = "astro-ph",
    doi = "10.1086/309216",
    SLACcitation = "%%CITATION = ASTRO-PH/9911177;%%"
}

@article{Cusin:2017fwz,
    author = "Cusin, Giulia and Pitrou, Cyril and Uzan, Jean-Philippe",
    title = "{Anisotropy of the astrophysical gravitational wave background: Analytic expression of the angular power spectrum and correlation with cosmological observations}",
    eprint = "1704.06184",
    archivePrefix = "arXiv",
    primaryClass = "astro-ph.CO",
    doi = "10.1103/PhysRevD.96.103019",
    journal = "Phys. Rev. D",
    volume = "96",
    number = "10",
    pages = "103019",
    year = "2017"
}

@article{Bartolo:2019oiq,
    author = "Bartolo, N. and Bertacca, D. and Matarrese, S. and Peloso, M. and Ricciardone, A. and Riotto, A. and Tasinato, G.",
    title = "{Anisotropies and non-Gaussianity of the Cosmological Gravitational Wave Background}",
    eprint = "1908.00527",
    archivePrefix = "arXiv",
    primaryClass = "astro-ph.CO",
    doi = "10.1103/PhysRevD.100.121501",
    journal = "Phys. Rev. D",
    volume = "100",
    number = "12",
    pages = "121501",
    year = "2019"
}

@ARTICLE{1996ApJ...469..437S,
       author = {{Seljak}, Uros and {Zaldarriaga}, Matias},
        title = "{A Line-of-Sight Integration Approach to Cosmic Microwave Background Anisotropies}",
      journal = {\apj},
         year = 1996,
        month = oct,
       volume = {469},
        pages = {437},
          doi = {10.1086/177793},
archivePrefix = {arXiv},
       eprint = {astro-ph/9603033},
 primaryClass = {astro-ph},
       adsurl = {https://ui.adsabs.harvard.edu/abs/1996ApJ...469..437S}
}

@ARTICLE{1985SvAL...11..271K,
       author = {{Kofman}, L.~A. and {Starobinskii}, A.~A.},
        title = "{Effect of the Cosmological Constant on Largescale Anisotropies in the Microwave Background}",
      journal = {Soviet Astronomy Letters},
         year = 1985,
        month = sep,
       volume = {11},
        pages = {271-274},
       adsurl = {https://ui.adsabs.harvard.edu/abs/1985SvAL...11..271K}
}

@ARTICLE{1967ApJ...147...73S,
       author = {{Sachs}, R.~K. and {Wolfe}, A.~M.},
        title = "{Perturbations of a Cosmological Model and Angular Variations of the Microwave Background}",
      journal = {\apj},
         year = 1967,
        month = jan,
       volume = {147},
        pages = {73},
          doi = {10.1086/148982},
       adsurl = {https://ui.adsabs.harvard.edu/abs/1967ApJ...147...73S}
}

@book{Dodelson:2003ft,
    author = "Dodelson, Scott",
    title = "{Modern Cosmology}",
    isbn = "978-0-12-219141-1",
    publisher = "Academic Press",
    address = "Amsterdam",
    year = "2003"
}

@article{Contaldi:2016koz,
    author = "Contaldi, Carlo R.",
    title = "{Anisotropies of Gravitational Wave Backgrounds: A Line Of Sight Approach}",
    eprint = "1609.08168",
    archivePrefix = "arXiv",
    primaryClass = "astro-ph.CO",
    reportNumber = "IMPERIAL-TP-2016-CC-2",
    doi = "10.1016/j.physletb.2017.05.020",
    journal = "Phys. Lett. B",
    volume = "771",
    pages = "9--12",
    year = "2017"
}

@article{LISACosmologyWorkingGroup:2022kbp,
    author = "Bartolo, Nicola and others",
    collaboration = "LISA Cosmology Working Group",
    title = "{Probing anisotropies of the Stochastic Gravitational Wave Background with LISA}",
    eprint = "2201.08782",
    archivePrefix = "arXiv",
    primaryClass = "astro-ph.CO",
    doi = "10.1088/1475-7516/2022/11/009",
    journal = "JCAP",
    volume = "11",
    pages = "009",
    year = "2022",
    number = " - "
}

@article{Mentasti:2023uyi,
    author = "Mentasti, Giorgio and Contaldi, Carlo R. and Peloso, Marco",
    title = "{Probing the galactic and extragalactic gravitational wave backgrounds with space-based interferometers}",
    eprint = "2312.10792",
    archivePrefix = "arXiv",
    primaryClass = "gr-qc",
    doi = "10.1088/1475-7516/2024/06/055",
    journal = "JCAP",
    volume = "06",
    pages = "055",
    year = "2024",
    number = " - "
}

@article{Mentasti:2025ywl,
    author = "Mentasti, Giorgio and Contaldi, Carlo R. and Peloso, Marco",
    title = "{Strong scale-dependence does not enhance the kinematic boosting of gravitational wave backgrounds}",
    eprint = "2507.16901",
    archivePrefix = "arXiv",
    primaryClass = "astro-ph.CO",
    reportNumber = "Imperial--TP--2025--CC--5, Imperial-TP-2025-CC-5",
    doi = "10.1088/1475-7516/2026/02/068",
    journal = "JCAP",
    volume = "02",
    pages = "068",
    year = "2026",
    number = " - "
}

@article{Pi:2020otn,
    author = "Pi, Shi and Sasaki, Misao",
    title = "{Gravitational Waves Induced by Scalar Perturbations with a Lognormal Peak}",
    eprint = "2005.12306",
    archivePrefix = "arXiv",
    primaryClass = "gr-qc",
    reportNumber = "YITP-20-75, YITP-75, IPMU20-0054",
    doi = "10.1088/1475-7516/2020/09/037",
    journal = "JCAP",
    volume = "09",
    pages = "037",
    year = "2020",
    number = " - "
}

@article{Heisenberg:2024var,
    author = "Heisenberg, Lavinia and Inchausp{\'e}, Henri and Maibach, David",
    title = "{Observing kinematic anisotropies of the stochastic background with LISA}",
    eprint = "2401.14849",
    archivePrefix = "arXiv",
    primaryClass = "gr-qc",
    doi = "10.1088/1475-7516/2025/01/044",
    journal = "JCAP",
    volume = "01",
    pages = "044",
    year = "2025",
    number = " - "
}

@article{Bartolo_2019,
   title={Primordial Black Hole Dark Matter: LISA Serendipity},
   volume={122},
   ISSN={1079-7114},
   url={http://dx.doi.org/10.1103/PhysRevLett.122.211301},
   DOI={10.1103/physrevlett.122.211301},
   number={21},
   journal={Physical Review Letters},
   publisher={American Physical Society (APS)},
   author={Bartolo, N. and De Luca, V. and Franciolini, G. and Lewis, A. and Peloso, M. and Riotto, A.},
   year={2019},
   month=may }

@article{Cusin:2022cbb,
    author = "Cusin, Giulia and Tasinato, Gianmassimo",
    title = "{Doppler boosting the stochastic gravitational wave background}",
    eprint = "2201.10464",
    archivePrefix = "arXiv",
    primaryClass = "astro-ph.CO",
    doi = "10.1088/1475-7516/2022/08/036",
    journal = "JCAP",
    volume = "08",
    number = "08",
    pages = "036",
    year = "2022"
}

@article{refId0,
	author = {{Planck Collaboration} and {Ade, P. A. R.} et al.},
	title = {Planck 2015 results - XXI. The integrated Sachs-Wolfe effect},
	DOI= "10.1051/0004-6361/201525831",
	url= "https://doi.org/10.1051/0004-6361/201525831",
	journal = {A\&A},
	year = 2016,
	volume = 594,
	pages = "A21",
}

@article{Baker:2019nia,
    author = "Baker, John and others",
    title = "{The Laser Interferometer Space Antenna: Unveiling the Millihertz Gravitational Wave Sky}",
    eprint = "1907.06482",
    archivePrefix = "arXiv",
    primaryClass = "astro-ph.IM",
    reportNumber = "FERMILAB-PUB-19-436-A",
    month = "7",
    year = "2019",
    journal = "-"
}

@article{doi:10.1142/S0217751X2050075X,
author = {Ruan, Wen-Hong and Guo, Zong-Kuan and Cai, Rong-Gen and Zhang, Yuan-Zhong},
title = {Taiji program: Gravitational-wave sources},
journal = {International Journal of Modern Physics A},
volume = {35},
number = {17},
pages = {2050075},
year = {2020},
doi = {10.1142/S0217751X2050075X},
URL = {https://doi.org/10.1142/S0217751X2050075X},
eprint = {https://doi.org/10.1142/S0217751X2050075X}
}

@article{Liang:2021bde,
    author = "Liang, Zheng-Cheng and Hu, Yi-Ming and Jiang, Yun and Cheng, Jun and Zhang, Jian-dong and Mei, Jianwei",
    title = "{Science with the TianQin Observatory: Preliminary results on stochastic gravitational-wave background}",
    eprint = "2107.08643",
    archivePrefix = "arXiv",
    primaryClass = "astro-ph.CO",
    doi = "10.1103/PhysRevD.105.022001",
    journal = "Phys. Rev. D",
    volume = "105",
    number = "2",
    pages = "022001",
    year = "2022"
}

@article{Caprini:2024hue,
    author = "Caprini, Chiara and Jinno, Ryusuke and Lewicki, Marek and Madge, Eric and Merchand, Marco and Nardini, Germano and Pieroni, Mauro and Roper Pol, Alberto and Vaskonen, Ville",
    collaboration = "LISA Cosmology Working Group",
    title = "{Gravitational waves from first-order phase transitions in LISA: reconstruction pipeline and physics interpretation}",
    eprint = "2403.03723",
    archivePrefix = "arXiv",
    primaryClass = "astro-ph.CO",
    reportNumber = "LISA-COSWG-24-01, CERN-TH-2024-029",
    doi = "10.1088/1475-7516/2024/10/020",
    journal = "JCAP",
    volume = "10",
    pages = "020",
    year = "2024",
    number = "-"
}

@article{Sasaki:2018dmp,
    author = "Sasaki, Misao and Suyama, Teruaki and Tanaka, Takahiro and Yokoyama, Shuichiro",
    title = "{Primordial black holes{\textemdash}perspectives in gravitational wave astronomy}",
    eprint = "1801.05235",
    archivePrefix = "arXiv",
    primaryClass = "astro-ph.CO",
    doi = "10.1088/1361-6382/aaa7b4",
    journal = "Class. Quant. Grav.",
    volume = "35",
    number = "6",
    pages = "063001",
    year = "2018"
}

@article{Damour:2004kw,
    author = "Damour, Thibault and Vilenkin, Alexander",
    title = "{Gravitational radiation from cosmic (super)strings: Bursts, stochastic background, and observational windows}",
    eprint = "hep-th/0410222",
    archivePrefix = "arXiv",
    doi = "10.1103/PhysRevD.71.063510",
    journal = "Phys. Rev. D",
    volume = "71",
    pages = "063510",
    year = "2005"
}

@article{baghiUncoveringGravitationalwaveBackgrounds2023,
  title = {Uncovering Gravitational-Wave Backgrounds from Noises of Unknown Shape with {{LISA}}},
  author = {Baghi, Quentin and Karnesis, Nikolaos and Bayle, Jean-Baptiste and Besan{\c c}on, Marc and Inchausp{\'e}, Henri},
  year = 2023,
  month = apr,
  journal = {Journal of Cosmology and Astroparticle Physics},
  volume = {2023},
  number = {04},
  pages = {066},
  publisher = {IOP Publishing},
  issn = {1475-7516},
  doi = {10.1088/1475-7516/2023/04/066}
}

@article{boileauSpectralSeparationStochastic2021a,
  title = {Spectral Separation of the Stochastic Gravitational-Wave Background for {{LISA}}: {{Observing}} Both Cosmological and Astrophysical Backgrounds},
  shorttitle = {Spectral Separation of the Stochastic Gravitational-Wave Background for {{LISA}}},
  author = {Boileau, Guillaume and Christensen, Nelson and Meyer, Renate and Cornish, Neil J.},
  year = 2021,
  month = may,
  journal = {Physical Review D},
  volume = {103},
  number = {10},
  pages = {103529},
  publisher = {American Physical Society},
  doi = {10.1103/PhysRevD.103.103529},
  urldate = {2025-11-18}
}

@misc{buscicchioFirstYearLISA2025,
  title = {The First Year of {{LISA Galactic}} Foreground},
  author = {Buscicchio, Riccardo and Pozzoli, Federico and Chirico, Daniele and Sesana, Alberto},
  year = 2025,
  month = nov,
  number = {arXiv:2511.03604},
  eprint = {2511.03604},
  primaryclass = {astro-ph},
  publisher = {arXiv},
  doi = {10.48550/arXiv.2511.03604},
  urldate = {2025-11-14},
  archiveprefix = {arXiv}
}

@article{christensenStochasticGravitationalWave2019,
  title = {Stochastic Gravitational Wave Backgrounds},
  author = {Christensen, Nelson},
  year = 2019,
  month = jan,
  journal = {Reports on Progress in Physics},
  volume = {82},
  number = {1},
  pages = {016903--016903},
  publisher = {Institute of Physics Publishing},
  doi = {10.1088/1361-6633/aae6b5}
}

@article{cornishDetectingStochasticGravitational2001,
  title = {Detecting a Stochastic Gravitational Wave Background with the {{Laser Interferometer Space Antenna}}},
  author = {Cornish, Neil J.},
  year = 2001,
  month = dec,
  journal = {Physical Review D},
  volume = {65},
  number = {2},
  pages = {022004},
  publisher = {American Physical Society},
  doi = {10.1103/PhysRevD.65.022004},
  urldate = {2026-06-05}
}

@misc{criswellTemplatedAnisotropicAnalyses2024,
  title = {Templated {{Anisotropic Analyses}} of the {{LISA Galactic Foreground}}},
  author = {Criswell, Alexander W. and Rieck, Steven and Mandic, Vuk},
  year = 2024,
  month = oct,
  number = {arXiv:2410.23260},
  eprint = {2410.23260},
  publisher = {arXiv},
  doi = {10.48550/arXiv.2410.23260},
  urldate = {2024-11-08},
  archiveprefix = {arXiv}
}

@article{capriniReconstructingSpectralShape2019,
  title = {Reconstructing the Spectral Shape of a Stochastic Gravitational Wave Background with {{LISA}}},
  author = {Caprini, Chiara and Figueroa, Daniel G. and Flauger, Raphael and Nardini, Germano and Peloso, Marco and Pieroni, Mauro and Ricciardone, Angelo and Tasinato, Gianmassimo},
  year = 2019,
  month = nov,
  journal = {Journal of Cosmology and Astroparticle Physics},
  volume = {2019},
  number = {11},
  pages = {017},
  publisher = {IOP Publishing},
  issn = {1475-7516},
  doi = {10.1088/1475-7516/2019/11/017},
  urldate = {2022-09-06},
  langid = {english}
}

@article{flaugerImprovedReconstructionStochastic2021,
  title = {Improved Reconstruction of a Stochastic Gravitational Wave Background with {{LISA}}},
  author = {Flauger, Raphael and Karnesis, Nikolaos and Nardini, Germano and Pieroni, Mauro and Ricciardone, Angelo and Torrado, Jes{\'u}s},
  year = 2021,
  month = jan,
  journal = {Journal of Cosmology and Astroparticle Physics},
  volume = {2021},
  number = {01},
  pages = {059},
  issn = {1475-7516},
  doi = {10.1088/1475-7516/2021/01/059},
  urldate = {2025-11-18},
  langid = {english}
}

@article{pozzoliWeaklyParametricApproach2024,
  title = {Weakly Parametric Approach to Stochastic Background Inference in {{LISA}}},
  author = {Pozzoli, Federico and Buscicchio, Riccardo and Moore, Christopher J. and Haardt, Francesco and Sesana, Alberto},
  year = 2024,
  month = apr,
  journal = {Physical Review D},
  volume = {109},
  number = {8},
  pages = {083029},
  publisher = {American Physical Society},
  doi = {10.1103/PhysRevD.109.083029},
  urldate = {2024-06-26}
}

@misc{santiniFlexibleGPUacceleratedApproach2025a,
  title = {A Flexible, {{GPU-accelerated}} Approach for the Joint Characterization of {{LISA}} Instrumental Noise and {{Stochastic Gravitational Wave Backgrounds}}},
  author = {Santini, Alessandro and Muratore, Martina and {Jonathan Gair} and Hartwig, Olaf},
  year = 2025,
  month = jul,
  number = {arXiv:2507.06300},
  eprint = {2507.06300},
  primaryclass = {gr-qc},
  publisher = {arXiv},
  doi = {10.48550/arXiv.2507.06300},
  urldate = {2025-09-01},
  archiveprefix = {arXiv}
}

@article{Foreman_Mackey_2013,
   title={<tt>emcee</tt>: The MCMC Hammer},
   volume={125},
   ISSN={1538-3873},
   url={http://dx.doi.org/10.1086/670067},
   DOI={10.1086/670067},
   number={925},
   journal={Publications of the Astronomical Society of the Pacific},
   publisher={IOP Publishing},
   author={Foreman-Mackey, Daniel and Hogg, David W. and Lang, Dustin and Goodman, Jonathan},
   year={2013},
   month=Mar, pages={306–312} }

@article{kumeAssessingImpactUnequal2025,
  title = {Assessing the {{Impact}} of {{Unequal Noises}} and {{Foreground Modeling}} on {{SGWB Reconstruction}} with {{LISA}}},
  author = {Kume, Jun'ya and Peloso, Marco and Pieroni, Mauro and Ricciardone, Angelo},
  year = 2025,
  month = jun,
  journal = {Journal of Cosmology and Astroparticle Physics},
  volume = {2025},
  number = {06},
  eprint = {2410.10342},
  primaryclass = {gr-qc},
  pages = {030},
  issn = {1475-7516},
  doi = {10.1088/1475-7516/2025/06/030},
  urldate = {2025-06-26},
  archiveprefix = {arXiv}
}

@article{chengDetectionStochasticGravitational2025,
  title = {Detection of the Stochastic Gravitational Wave Background with the Space-Based Gravitational-Wave Detector Network},
  author = {Cheng, Jun and Li, En-Kun and Mei, Jianwei},
  year = 2025,
  month = nov,
  journal = {Physics Letters B},
  volume = {870},
  pages = {139890},
  issn = {0370-2693},
  doi = {10.1016/j.physletb.2025.139890},
  urldate = {2026-06-05}
}

@article{liangUnveilingMulticomponentStochastic2025,
  title = {Unveiling a Multicomponent Stochastic Gravitational-Wave Background with the \$\textbackslash mathrm\textbraceleft{{TianQin}}\textbraceright +\textbackslash mathrm\textbraceleft{{LISA}}\textbraceright\$ Network},
  author = {Liang, Zheng-Cheng and Li, Zhi-Yuan and Li, En-Kun and Zhang, Jian-dong and Hu, Yi-Ming},
  year = 2025,
  month = feb,
  journal = {Physical Review D},
  volume = {111},
  number = {4},
  pages = {043032},
  publisher = {American Physical Society},
  doi = {10.1103/PhysRevD.111.043032},
  urldate = {2026-06-05}
}

@article{wangAbilityLISATaiji2024,
  title = {Ability of {{LISA}}, {{Taiji}}, and Their Networks to Detect the Stochastic Gravitational Wave Background Generated by Cosmic Strings},
  author = {Wang, Bo-Rui and Li, Jin},
  year = 2024,
  month = mar,
  journal = {Physical Review D},
  volume = {109},
  number = {6},
  pages = {063520},
  publisher = {American Physical Society},
  doi = {10.1103/PhysRevD.109.063520},
  urldate = {2026-06-05}
}

@article{princeLISAOptimalSensitivity2002,
  title = {{{LISA}} Optimal Sensitivity},
  author = {Prince, Thomas A. and Armstrong, J. W. and Tinto, Massimo and Larson, Shane L.},
  year = 2002,
  journal = {Physical Review D - Particles, Fields, Gravitation and Cosmology},
  volume = {66},
  number = {12},
  issn = {15502368},
  doi = {10.1103/PhysRevD.66.122002}
}

@article{Zonca2019healpy,
  doi = {10.21105/joss.01298},
  url = {https://doi.org/10.21105/joss.01298},
  year = {2019},
  publisher = {The Open Journal},
  volume = {4},
  number = {35},
  pages = {1298},
  author = {Andrea Zonca and Leo P. Singer and Daniel Lenz and Martin Reinecke and Cyrille Rosset and Eric Hivon and Krzysztof M. Górski},
  title = {healpy: equal area pixelization and spherical harmonics transforms for data on the sphere in Python},
  journal = {Journal of Open Source Software}
}

@article{Bartolo:2019yeu,
    author = "Bartolo, Nicola and Bertacca, Daniele and Matarrese, Sabino and Peloso, Marco and Ricciardone, Angelo and Riotto, Antonio and Tasinato, Gianmassimo",
    title = "{Characterizing the cosmological gravitational wave background: Anisotropies and non-Gaussianity}",
    eprint = "1912.09433",
    archivePrefix = "arXiv",
    primaryClass = "astro-ph.CO",
    doi = "10.1103/PhysRevD.102.023527",
    journal = "Phys. Rev. D",
    volume = "102",
    number = "2",
    pages = "023527",
    year = "2020"
}

@article{Bravo:2025csu,
    author = "Bravo, Rafael and Riquelme, Walter",
    title = "{Imprints of Large-Scale Structures in the Anisotropies of the Cosmological Gravitational Wave Background}",
    eprint = "2505.15084",
    archivePrefix = "arXiv",
    primaryClass = "astro-ph.CO",
    month = "5",
    year = "2025",
    journal="-"
}

@article{Zhao:2024gan,
    author = "Zhao, Zhi-Chao and Wang, Sai and Li, Jun-Peng and Kohri, Kazunori",
    title = "{Study of primordial non-Gaussianity $f_{\textrm{NL}}$ and $g_{\textrm{NL}}$ with the cross-correlations between the scalar-induced gravitational waves and the cosmic microwave background}",
    eprint = "2412.02500",
    archivePrefix = "arXiv",
    primaryClass = "astro-ph.CO",
    reportNumber = "KEK-Cosmo-0364, KEK-TH-2670, KEK-QUP-2024-0025,
  https://github.com/Zhi-ChaoZhao/sigw{\_}class, KEK-QUP-2024-0025",
    doi = "10.1140/epjc/s10052-025-15115-8",
    journal = "Eur. Phys. J. C",
    volume = "85",
    number = "12",
    pages = "1406",
    year = "2025"
}

@article{Cai:2024dya,
    author = "Cai, Rong-Gen and Wang, Shao-Jiang and Yuwen, Zi-Yan and Zeng, Xiang-Xi",
    title = "{Anisotropies of cosmological gravitational wave backgrounds in non-flat spacetime}",
    eprint = "2410.17721",
    archivePrefix = "arXiv",
    primaryClass = "astro-ph.CO",
    doi = "10.1088/1475-7516/2025/01/011",
    journal = "JCAP",
    volume = "01",
    pages = "011",
    year = "2025",
    number = "-"
}

\end{document}